# Role of microstructure and structural disorder on tribological properties of polycrystalline diamond films

P. K. Ajikumar,[1,†,‡] K. Ganesan,[1,†,‡] N. Kumar,[1,†,‡] T. R. Ravindran,[2] S. Kalavathi[2] and M. Kamruddin[3]

[1.] Surface & Nanoscience Division, Materials Science Group, Indira Gandhi Centre for Atomic Research, Kalpakkam, India
[2.] Condensed Matter Physics Division, Materials Science Group, Indira Gandhi Centre for Atomic Research, Kalpakkam, India
[3.] Accelerator & Nanoscience Group, Materials Science Group, Indira Gandhi Centre for Atomic Research, Kalpakkam, India

**Abstract**

Polycrystalline diamond films with systematic change in microstructure that varies from microcrystalline to nanocrystalline structure are synthesized on Si by hot filament chemical vapor deposition. The morphology and structural properties of the grown diamond films are analyzed using field emission scanning electron microscope (FESEM), atomic force microscope (AFM), X-ray diffraction and Raman spectroscopy. The average roughness and grain size of the diamond films decrease with increase in $CH_4$ to $H_2$ ratio from 0.5 to 3 %. Also, structural disorder in these diamond films increases with decrease in grain size as evidenced from Raman spectroscopy. The coefficient of friction (CoF) is found to be very low for all the films. However, the average CoF is found to increase from 0.011±0.005 to 0.03±0.015 as the grain size decrease from ∼ 1 μm down to ∼20 nm. Post analysis of wear track by FESEM, AFM based nanoscale friction and Raman spectroscopy reveal that microcrystalline diamond undergoes shear induced amorphization with negligible wear rate while nanocrystalline diamond films undergo shear induced plastic deformation without amorphization. A comprehensive mechanism for the observed CoF is discussed in the framework of microstructure, structural disorder and shear induced tribo-chemical reactions at the sliding interface.

[†] Both authors contributed equally to this work.
[‡] Corresponding authors
Email : aji@igcar.gov.in (P.K.A), kganesan@igcar.gov.in (K.G), niranjan@igcar.gov.in (N.K)

# 1. Introduction

Diamond possesses some of the most useful physical properties such as high hardness, wear resistance, optical transparency, thermal conductivity, unique semiconductor characteristics, lowest compressibility, intrinsically low friction and inertness towards most chemical reagents. These unique characteristics of diamond make it attractive for several applications including wear resistant coatings, cutting tools for non-ferrous materials, as heat spreaders for electronic devices, flat panel displays and transparent windows for infrared light [1,2,3,4]. However, the practical use of diamonds in science and engineering applications are limited due to their scarcity and high cost. During the past few decades, a lot of research is being carried out for harnessing the full potential of diamond after successful development of synthetic diamonds by chemical vapor deposition (CVD). Diamond can be synthesized by a variety of CVD processes, including hot filament (HF), microwave plasma, DC plasma, plasma jet, arc discharge and combustion flame methods [1]. Among these methods, HFCVD is one of the most popular techniques because of its simplicity, low capital cost and the ability to scale up.

During HFCVD process, various gas phase radicals are produced on and near the hot filament and the chemistry of these radicals on the substrate plays an important role in the growth of diamond film [5,6]. Conventionally, HFCVD grown microcrystalline diamond (MCD) films exhibit a faceted morphology with large surface roughness when it is grown with low $CH_4$ to $H_2$ ratio. But, the surface morphology of the diamond films can be made smooth by increasing $CH_4$ to $H_2$ ratio that allows formation of more secondary nucleation which results in decrease of grain size leading to nanocrystalline diamond (NCD) [1]. However, NCD films consist of a large amount of $sp^2$-rich graphitic and amorphous carbon in grain boundaries which increase the structural disorder in the films. Moreover, the small grain size enhances toughness, without much compromise on mechanical properties, of NCD films. Such NCD films with high hardness and high toughness is of great interest in micro-/nano- electromechanical systems [7,8]. The reported hardness value for MCD is found to vary from 60 upto 100 GPa [9]. However, hardness is lower for NCD films and it varies upto 65 GPa depending upon the diamond structural characteristics [10].

The microstructure and surface roughness play an important role in several applications. For example, highly oriented and large grain size MCD film performs better than its NCD counterpart as heat sink in several applications including electronic devices [3]. MCD with smooth surface is more preferred in electronic and optical components whereas NCD films with a rough surface and high $sp^2$-rich grain boundaries perform better in field emitting devices [11]. Thus, the MCD and NCD films have their own specific characteristics that allow one to choose either of the material depending upon the requirement. To this end, CVD process parameters are varied to tune microstructure and structural disorder in the diamond films.

Diamond is an indispensable material in mechanical and tribological applications because of its combined characteristics of extreme hardness and ultralow friction. Typically, the coefficient of friction (CoF) of polycrystalline diamond films ranges between high to ultralow value depending upon tribo test conditions [12-25]. Some of the parameters affecting the tribological properties of diamond are contact pressure, crystallographic orientation, counterbody and surface roughness of the film [13,14,15,16]. More importantly, tribofilm formation is one of the critical mechanisms to explain the ultralow friction and wear on diamond films [2,17,18]. Apart from the above, recent studies demonstrate the importance of surface chemistry and tribological environment on the nature of friction and wear [8,12,188,19]. For example, in a humid atmosphere, friction and wear of diamond and tetrahedral amorphous carbon films are significantly reduced due to surface passivation and chemical rehybridization of $sp^3$ to $sp^2$ bonding [8,12,18,20,21]. Further, CoF on diamond coatings reduces with decrease in surface roughness as reported by several groups [22,23,24]. The rough surfaces introduce additional mechanical resistance due to deformation. However, Schneider *et al.* [25] had shown that smoothing the surface of diamond does not decrease friction coefficient, rather increases friction coefficient marginally. Hence, it is noted that despite having a lot of reports on the ultralow friction and wear behavior of diamond, the mechanism for CoF is not clearly understood in polycrystalline diamond film, especially the correlation with grain size and their structural properties. These facts have motivated the current investigations on the role of microstructure and structural disorder on tribological properties of polycrystalline diamond films.

Diamond films with tailored microstructure and structural disorder are grown in a custom designed HFCVD system by suitably controlling the $CH_4$ to $H_2$ feedstock gas ratios. The microstructure and structural disorder of these films are thoroughly studied by Field Emission Scanning Electron Microscopy (FESEM), Atomic Force Microscopy (AFM), X-ray diffraction (XRD) and Raman spectroscopy. The influence of microstructure and structural disorder of the diamond films on the tribological properties are analyzed. Frictional force microscopy (FFM) and Raman spectroscopy are performed inside the wear track to study the formation of tribofilm and its chemical structure on diamond surface. Based on these results, a plausible mechanism for the observed CoF of diamonds is discussed in the framework of microstructure and structural disorder of the diamond films.

**2. Experimental**

Diamond films were grown by a custom designed HFCVD system fabricated with cylindrical vacuum chamber of 20 and 60 cm inner diameter and height, respectively. Prior to growth, the reactor chamber was evacuated to a base vacuum of $2 \times 10^{-6}$ mbar using a turbo molecular pump. Then, the tungsten filament was carburized at ~ 1500 $^0$C for its stability during growth. The commercially polished Si (111) wafer of dimension 40 x 20 mm$^2$ was chemo-mechanically polished with micron level diamond paste to enhance diamond nucleation and was loaded into the reactor chamber. Maintaining a constant substrate temperature distribution is very important to control the uniformity of diamond coating. Consequently, an additional resistive heater is used to control the uniform substrate temperature with an accuracy of ± 1 $^0$C using a PID controller. In addition, the space between the adjacent tungsten filaments and distance from the tungsten filament to substrate are optimized to provide constant temperature and carbon radicals distribution over the entire substrate for uniform diamond growth. The substrate was placed at ~ 10 mm below the filament and temperature was raised to 800ºC. The filament temperature was monitored by an optical pyrometer and maintained at ~ 2000 ºC. Methane and hydrogen gases were used as source and dilution gases, respectively. The methane to hydrogen ratio, $CH_4 / CH_4+H_2$, was varied from 0.5 to 3% to achieve diamond with various morphologies and structural disorders. The total gas flow of $CH_4$ and $H_2$ was maintained at 100 sccm for all experiments through mass flow controllers. The operating pressure was maintained at 30 mbar throughout the synthesis using a needle valve. Growth time was fixed at 5 hours for all experiments. Uniform diamond films with superior adhesion were obtained over the entire area of the Si wafer. The thickness of the diamond films is in the range of 1 – 2 µm depending upon the growth process parameters.

Phase purity of the diamond films was evaluated by X-ray diffraction analysis in glancing incidence mode using Cu-K$_\alpha$ radiation (STOE, X-ray powder diffractometer). Linear reciprocating mode of a ball on disk tribometer (CSM, Switzerland) was used to carry out the tribological tests of the films. A spherical alumina ball of 6 mm diameter with a hardness of 12 GPa was used as a sliding body to measure the value of CoF. Alumina was chosen as the counter pair mainly due to its high hardness, oxidation resistance and chemical inertness. This would by and large avoid oxidation wear and provide effective tribological response from diamond film. The normal load and sliding speed were kept constant at moderate values of 2 N and 5 cm/s, respectively. Tests were performed at ambient (dry and unlubricated) conditions and room temperature. A FESEM (Supra 55, Carl Zeiss, Germany) was used to examine the variation in microstructure and morphology of as grown surface and wear tracks of the samples. Raman spectra were recorded using a micro-Raman spectrometer (Invia, Renishaw, UK) on the surface and wear track of the films using an argon ion laser with 514.5 nm and a grating with 1800 grooves/mm. A microscope objective of 100X magnification with numerical aperture of 0.8 was used in this study and the probing laser beam diameter was ~ 1 µm. Surface topography of the films were measured using an AFM (NTGRA, NT-MDT, Russia) in semi-contact mode. For nanoscale friction analysis on wear track, the topography and frictional force images were measured simultaneously in contact mode with an applied normal load of 20 nN using frictional force microscopy (FFM). The cantilever axis was maintained perpendicular to the scan direction in order to monitor the lateral deflection of the cantilever.

## 3. Results

Figure 1 shows the typical XRD pattern of the synthesized diamond films. The observed peaks match well with the cubic diamond (ICDD card no. C-00-006-0675). Inset of the pattern shows broadening of the XRD peak ($2\theta=44^0$) as a function of increasing CH$_4$ concentration in the feedstock gas. The increase in peak broadening implies either reduction in grain size or increase in strain with CH$_4$ concentration. Phase purity of the diamond is evident from the absence of any additional peak other than (111) and (220) planes of cubic diamond in the measured 2θ range.

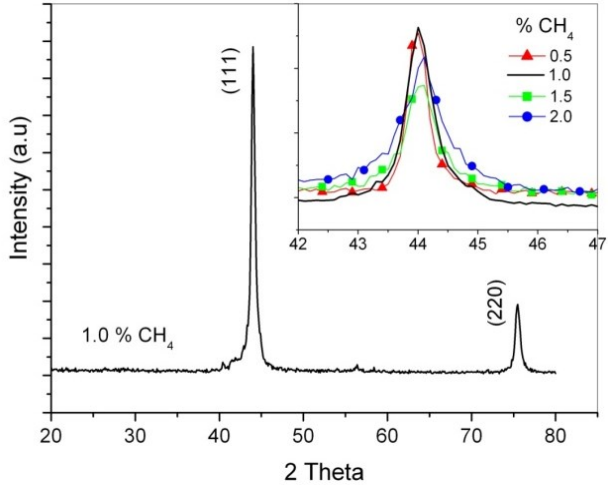

Fig. 1. Typical GIXRD pattern of diamond film. Inset shows broadening of (111) diffraction peak as a function of CH$_4$ concentration in the feedstock.

Figure 2 shows the FESEM micrographs of diamond films grown with different CH$_4$ concentrations in feedstock gas. The surface morphologies of the films are clearly distinct. Figure 2a corresponds to faceted MCD synthesized with 0.5% of CH$_4$ and it depicts a <110> oriented microcrystalline structure which creates fivefold symmetry [26]. The occurrence of fivefold symmetry is due to multiple twinning in {111} planes. These twin boundaries are completely free from dislocations and thus the misfit arising out of fivefold symmetry is accommodated by elastic strain alone [26]. The surface morphology of specimen prepared at 1.0 % of CH$_4$ (Fig. 2b) shows a combination of MCD with a large amount of NCD as secondary nucleation on all facets of crystals. Figure 2c, 2d and 2e, which correspond to 1.5, 2.0 and 3.0 % of CH$_4$ concentrations respectively, show considerable reduction in grain size with increase in CH$_4$ concentration. As a consequence of increased CH$_4$ content, nucleation density of diamond crystallites increases and it results in reduction of grain size. The grain size of MCD film was directly measured from SEM as well as AFM measurements. However, an accurate estimation of grain size using SEM and AFM was difficult on the diamond films with mixed MCD and NCD crystallites ( CH$_4$ ≥ 1%) and NCD crystallites ( CH$_4$ =3%) because of the limitation of these techniques. Hence, the crystallite size was estimated from XRD pattern using Scherrer formula.

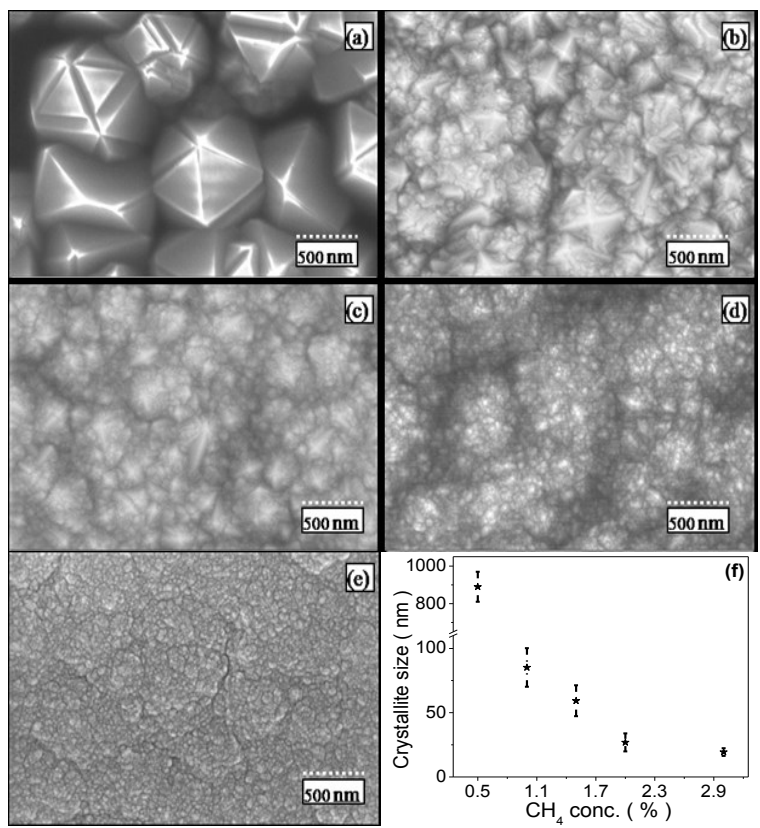

Fig. 2. FESEM micrographs of diamond films synthesized at different $CH_4$ concentrations, (a) 0.5 %, (b) 1.0 %, (c) 1.5 %, (d) 2.0 % and (e) 3.0 %. (f) The variation of crystallite size as a function of $CH_4$ concentration in the feedstock, measured by FESEM and X-ray diffraction.

Thus, the average crystalline size is found to be ~ 890 ± 80, 85.2 ± 15.4, 59.3 ± 12.4, 36.9 ± 7.1 and 19.4 ± 3.2 nm for the diamond films grown with increasing methane concentration of 0.5, 1.0, 1.5, 2 and 3 %, respectively (Fig. 2f). In order to measure the surface roughness, AFM topography analysis was performed as discussed in the next paragraph.

Figure 3 displays the AFM surface topography of the diamond films grown at different $CH_4$ concentrations. As discussed earlier, the topography is almost similar to the morphology observed by FESEM. The crystallite grain size decreases monotonically from ~ 1 μm down to ~ 20 nm with increase in $CH_4$ concentration in the feedstock as can be evidenced from Fig.3. However, the topography (Fig. 3d) appears to be agglomerated due to the AFM tip-sample convolution effect on NCD films. The root mean square (rms) roughness of the films is found to be 70 ± 8, 51±7, 35±5, 20±4 and 16±7 nm for the diamond films grown with $CH_4$ concentration of 0.5, 1.0, 1.5, 2.0 and 3.0 % respectively, measured over an area of 5 x 5 $\mu m^2$.

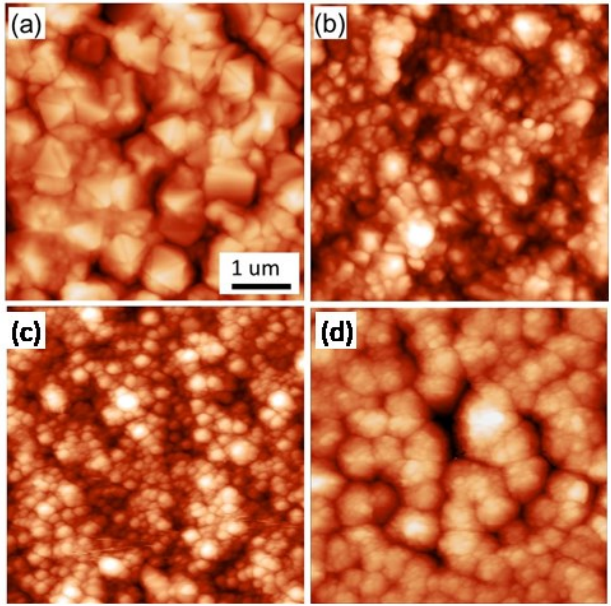

Fig.3. AFM topography of diamond films synthesized at different $CH_4$ concentrations, (a) 0.5 %, (b) 1.0 %, (c) 1.5 %, and (d) 3.0 %. The scale bar is same for all the images.

Figure 4 shows the Raman spectra of diamond films grown as a function of $CH_4$ concentration in the feedstock. The MCD film (0.5% of $CH_4$) shows a single sharp Raman band at ~1332 $cm^{-1}$. The absence of other Raman bands, especially in the range 1500–1550 $cm^{-1}$, confirms phase purity of diamond lattice [27]. Raman spectrum is in complete agreement with the FESEM micrograph (Fig.2a) which shows well-developed faceted diamond crystals. With increase in concentration of $CH_4$ in feedstock, several additional peaks are observed at ~ 1140, 1350, 1480 and 1550 $cm^{-1}$ in the Raman spectra (Fig. 4). The peaks at 1350 and 1550 $cm^{-1}$ correspond to D and G bands of disordered graphitic structures, respectively. Increase in intensity and broadening of the D-and G- bands indicate enhancement of defects and disorder in diamond phase [27]. However, the peak intensities alone cannot be directly correlated to the amount of diamond and graphitic phases since Raman scattering cross section of graphitic bonds is much higher ( ~ 250 times) than that of diamond [27]. The two additional peaks in the Raman spectra, at 1140 and 1480 cm$^{-1}$, is the signature of transpoly-acetylene structure and they arise due to the presence of H in the disordered carbon network mainly at grain boundaries [27,28]. Overall, the Raman spectroscopy clearly indicates that the structural disorder increases monotonically with increasing $CH_4$ concentration in the process gas.

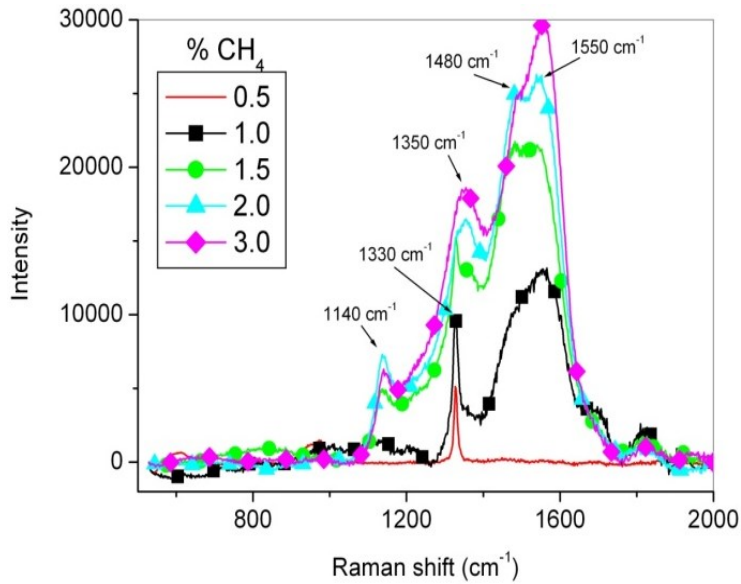

Figure 4. Raman spectra of the diamond films grown at different $CH_4$ concentrations.

Figure 5a shows the CoF curves as a function of sliding distance for the diamond films against alumina counterbody. In general, CoF exhibits a similar trend for all these diamond films. At initial sliding, the CoF is high (upto ~ 0.55) corresponding to static friction with a short run-in regime, which is not exceeding a sliding distance of 10 m. This initial high friction results from an intense mechanical interlocking between hard surface asperities of the interfaces. After the run-in, a steady state regime is reached with low friction for the remaining sliding distance of the tribotest. It should be noted that the run-in distance for MCD is very small (~1.5 m) indicating the extremely low energy dissipation during tribotest. Further, the run-in distance slowly increases with decrease in grain size, in general, as shown in inset of Fig. 5a and the NCD film grown with 3% $CH_4$ has a run-in distance of ~ 9 m. Figure 5b shows the average CoF which is measured over three wear tracks. It is observed that MCD film with faceted morphology exhibits a minimum average CoF of ~ 0.011±0.005 whereas NCD film grown with 3.0 % $CH_4$ has maximum value (~ 0.033 ± 0.015 ) over the sliding distance of 100 m. The friction coefficient of all other diamond films is in between these two values. The formation of tribofilm or transfer film during wear test is considered one of the most important parameters in understanding the friction coefficient of carbon films. The presence of tribofilm and its chemical structure on diamond surface are evaluated by SEM, FFM and Raman spectroscopy and they are discussed in the forthcoming paragraphs.

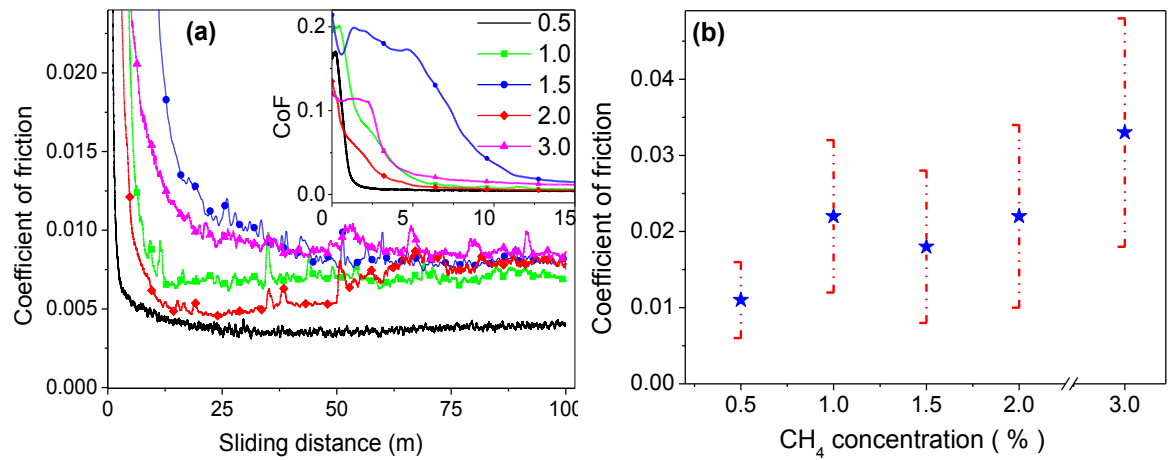

Fig.5(a) Typical friction coefficient curve as a function of sliding distance for the diamond films grown under different CH$_4$ concentration. The inset shows the magnified part of the run-in regime (b) The average friction coefficient of the diamond films for a sliding distance of 100 meters.

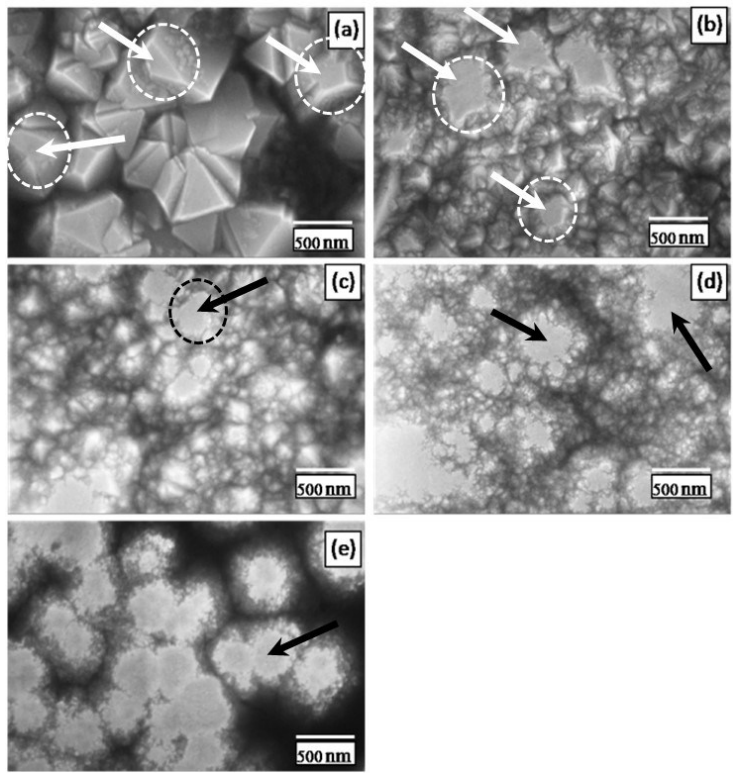

Figure 6. FESEM images of wear tracks showing the wear damage on diamond coatings synthesized at different methane concentrations, (a) 0.5 %, (b) 1.0 %, (c) 1.5 %, (d) 2.0 % and (e) 3.0 %. The arrow marks and circles represent the worn-out surface of diamond.

Figure 6 (a)-(e) shows the FESEM micrographs of tribo-tracks developed during one set of tribo measurements. As shown in Fig.6, the wear damage increases for the diamond films grown with increase in $CH_4$ concentration. The MCD film has the minimum wear damage wherein a few localized pyramidal surfaces are worn-out during the tribo tests (as shown by the arrow marks in Fig. 6a). It is clearly shown that wear induced deformation increases with decrease in grain size and the deformation is very high for NCD film deposited with 3% $CH_4$, as evidenced from Fig 6e. In other words, the diamond film with higher structural disorder having a large amount of grain boundaries, $sp^2$-rich graphitic and amorphous carbon undergo maximum wear damage. Moreover, the wear is also directly linked with surface roughness of the films. The roughness of MCD film is higher than NCD films and this is responsible for smaller contact interfaces.

Figure 7a and 7b show the topography and its corresponding FFM image of MCD film on the worn-out surface. Similar to SEM micrograph, the topography also shows flattened worn-out regions as indicated by circle and an arrow mark but it is not clear from this image whether tribofilm is present or not. However, the FFM image exhibits a small change in contrast, especially on worn-out regions marked by arrow marks have lower friction (dark regions) than the other grain boundaries and microstructures. Fig. 7c shows the topography of a part of the wear track for the film grown with 1.5% of $CH_4$. The worn-out region consists of thick transfer film with flat surface which is seen as bright region at middle of the image surrounded by grainy structure on both sides (Fig.7c). Fig. 7d & 7e are the magnified part of topography and FFM mappings respectively, from the marked area in Fig.7c. The FFM mapping (Fig. 7e) clearly shows that the transfer film has lower friction (dark regions) than the grainy structure.

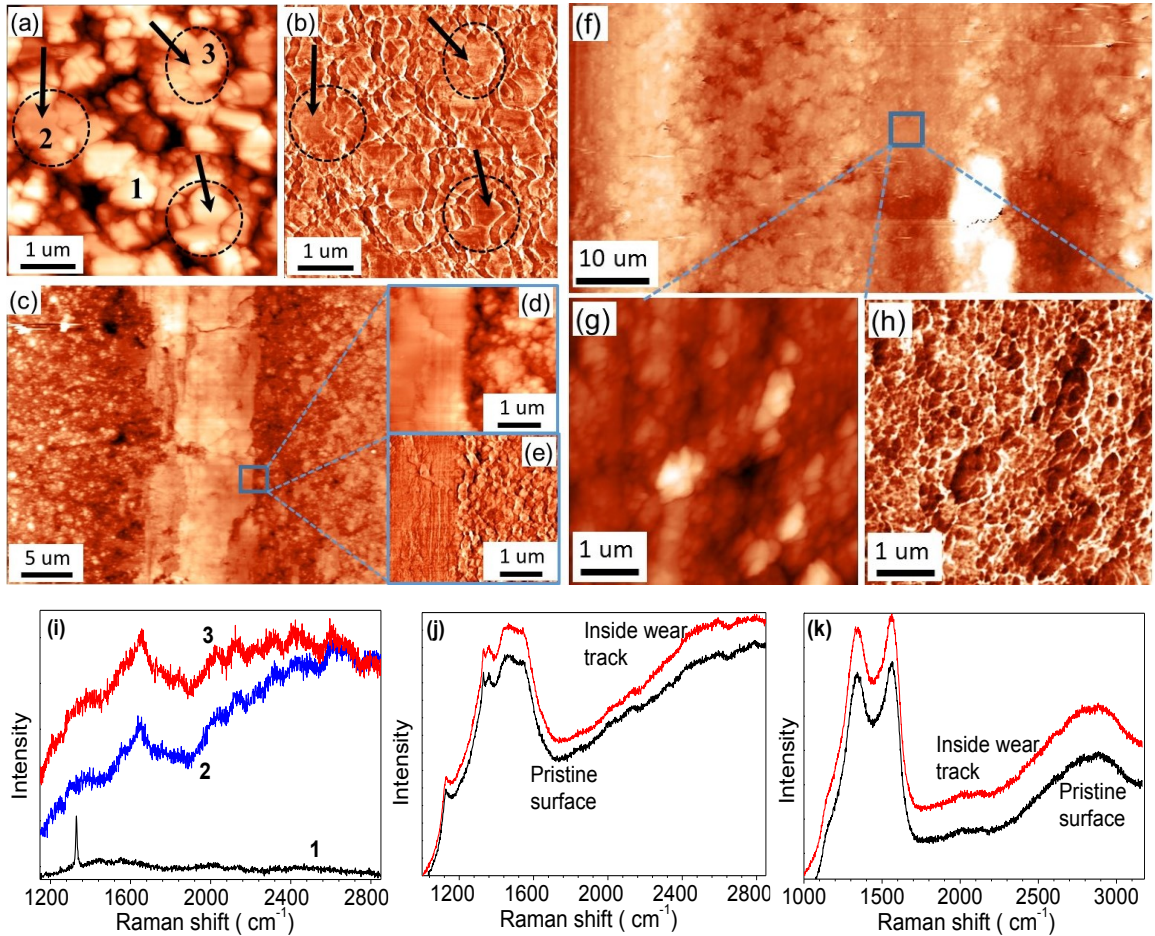

Fig.7. Simultaneously recorded (a) topography and (b) frictional force micrographs of microcrystalline diamond grown with 0.5% of CH$_4$, inside a wear track. Topography of diamond films measured inside a wear track on diamond films grown with (c) 1.5 and (f) 3 % of CH$_4$. A magnified part of (d) topography and (e) frictional force micrographs of diamond film (CH$_4$ = 1.5 %) from the selected location on Fig 7c. A magnified part of (g) topography and (h) frictional force micrographs of diamond film (CH$_4$ = 3 %) from the selected location on Fig. 7f. (i) Raman spectra recorded inside the wear track at representative positions of 1, 2 and 3 as indicated in Fig.7a. Raman spectra recorded on pristine surface and inside wear track for the diamond films grown with (j) 1.5 and (k) 3 % of CH$_4$.

Figure 7f shows the topography on the worn-out surface of NCD film (3 % of CH$_4$) exhibiting very rough morphology with *rms* roughness of ~ 160 nm measured over an area of 40 x 80 um$^2$, which is much higher as compared to roughness of pristine surface. Also, the peak to peak height variation is found upto 4 μm which can occur due to the segregation of debris that forms during tribotest. In addition, a magnified part of topography (Fig.7g) on worn-out surface also shows a grainy structure even after removal of materials during tribotest. The very rough morphology and grainy structures inside the wear track indicate that a stable and adherent tribofilm is not formed on diamond surface. The corresponding FFM mapping indicates a low friction regime (dark region) on grain surfaces and high friction regime in and around grain boundaries as shown in Fig. 7h. Thus, the FFM measurement clearly reveal the presence of inhomogeneous distribution of friction within the wear track.

Figure 7i shows Raman spectra recorded inside wear track on representative positions 1, 2 and 3 as marked in Fig 7a. The numbers 1, 2 & 3 in Fig. 7a are only a representation of the location, not the actual location where the Raman spectra were collected. Also, the number 1 represents a pristine diamond surface, and 2 and 3 represent a worn out surface inside the wear track. In general, the most of the Raman spectra are similar to pristine MCD as represented in curve 1 (Fig 7i). However, Raman spectra on several spots do have a signature of "diamond-like" amorphous carbon as represented in curves 2 and 3 (Fig 7i). Based on statistical analysis, we attribute these Raman spectra to the presence of amorphous carbon which can arise due to wear induced amorphization of diamond during tribo measurements. In contrast, Raman spectra of NCD films, grown with 1.5 and 3 % of $CH_4$, do not significantly vary from the pristine and worn-out surfaces as can be evidenced from Fig 7j and Fig. 7k, respectively. This indicates that NCD films do not undergo tribo-chemical induced phase transition during wear test.

## 4. Discussion

Based on the above observations, we summarize the results as follows. A systematic variation in microstructure ranging from MCD to NCD is obtained with increase in $CH_4$ concentration in the feedstock during growth. The grain size and surface roughness of the diamond films decreases but their structural disorder increases with increase in $CH_4$ concentration. Although, all the diamond films have low CoF ~ < 0.03, the CoF have a trend to increase with $CH_4$ concentration. In other words, CoF increases with decrease in grain size which is contrary to common perception. Further, the wear in MCD is much smaller in comparison to NCD films. In addition, the tribofilm analysis performed inside the wear track reveals that the shear induced amorphization on MCD but there exists no such phase transition on NCD films.

In the present study, the observed behavior of CoF can be understood as follows: The actual area of tribo-contact is very low for MCD and thus the pressure under the contact points is extremely high which can convert the $sp^3$ diamond into $sp^2$-rich amorphous carbon via pressure induced amorphization [29]. Here, Raman spectra provide a clear evidence for the presence of amorphous carbon on the worn-out surface and it supports the pressure / shear induced amorphization mechanism of diamond. Furthermore, low friction in MCD is directly related to small contact interfaces during the tribotest and hence, energy dissipation decreases. Pastewka *et al.* [30] demonstrated by molecular dynamics study that diamond surface undergoes structural phase transition from $sp^3$ to $sp^2$ – rich amorphous structure when diamond faces are rubbed against one another. In addition, the formation of amorphous adlayer and wear rate on diamond also strongly depend on surface orientation and sliding direction during wear test. Here, if the shear induced amorphization processes were continuous during wear test, the wear rate would be large due to the removal of amorphous material via unstable CO and $CO_2$ formation under atmospheric wear conditions. However, the removal of material is not observed on MCD films except for a minor worn-out surface. Hence, it can be inferred that the amorphous adlayer undergoes surface passivation which prevents further wear and also decrease friction on diamond surface [188,20,31,32,33,34]. At ambient atmospheric conditions during wear test, the diamond surface can easily undergo –H and –OH termination due to the dissociative passivation of $H_2O$ molecules [34]. Especially, diamond or diamond-like carbon surface with –H termination shows superlubricity due to the electrostatic repulsion between the counter faces [8,35]. It is also noted here that the as grown diamond is always terminated with large amount of –H which can also participate on reduction of CoF on diamond. Furthermore, Kuwahara et al [18] observe different regimes of friction on diamond (111) varying from very high to superlubricity as a function of water molecules in the tribo-envirnment. The most important observation is that an aromatic passivation of surface via Pandey surface reconstruction occurs with a sufficient amount of water molecules that can lead to formation of graphene nanostructures on diamond and amorphous carbon surfaces [188]. Hence, the observed low friction on MCD can be attributed to shear induced amorphization at a very short run-in distance and subsequently either surface passivation by -H / -OH molecules or the formation of graphene nanostructures through Pandey surface reconstruction or by both the mechanisms.

The CoF on all other diamond films grown at higher $CH_4$ concentration ( >1 %) have slightly higher than MCD. Also, these diamond films have structural disorder with random oriented grains, grain boundaries and large amount of $sp^2$-rich carbon bonding. During wear test, the contact areas between the two surfaces also increase with decrease in grain size. Thus, the contact pressure on these diamond surface is low during wear test that may not allow any shear induced phase transition. Raman spectra recorded inside and outside the wear track also look almost similar to each other confirming the absence of structural phase transition. However, the wear rate is found to increase with increase in structural disorder which can be attributed to the shear induced plastic deformation. The presence of disorder and large amount of dangling bonds make these $sp^2$-rich carbon bondings to be highly reactive to the environment and hence, the adhesion between the counterfaces becomes very high [8,12,36]. Consequently, the tractive force of adhesion and friction induced heat lead to the formation of spalling pits [12,36]. Further, the shed off diamond grains coalesce to form as tribolayer on NCD film surface. However, the diamond particles in the tribolayer do not bind themselves strong enough and hence, they are continuously removed from NCD surface. Thus, the NCD films undergo high wear rate and high friction due to the adhesion induced spalling pits on the NCD surface. In addition, the local topography and FFM mappings also confirm the high surface roughness (~ 160 nm ) on the wear track of NCD film. Further, the molecular dynamics study by Gao *et al.* [37] had also concluded that surfaces with a higher fraction of $sp^2$-bonded carbon are likely to exhibit higher friction due to the tribochemical reactions and formation of chemical bonds across the interface leading to atomic-scale wear events. Moreover, a systematic increase in graphitic carbon content and trans-polyacetylene (Fig.3) as evidenced by Raman spectroscopy also corroborate the increase in friction coefficient as a function of $CH_4$ concentration in the feedstock. Therefore, the hardness of diamond and chemical stability of tribolayer are the governing mechanisms for improved tribological properties of MCD.

## 5. Conclusions

Diamond films with different morphologies, varying from microcrystalline to nanocrystalline structure are successfully synthesized by varying $CH_4$ to $H_2$ ratio in the feedstock gas using hot filament chemical vapor deposition (HFCVD). The surface roughness and $sp^2$-rich graphitic carbon are found to decrease and increase, respectively with increase in $CH_4$ concentration in the feedstock. Ultralow CoF is measured for all the films in ambient condition. However, the CoF is found to be ~ 0.011±0.005 for microcrystalline diamond (MCD) while fine grained and relatively smooth nanocrystalline diamond (NCD) showed slightly higher CoF (~0.033±0.015). The shear induced amorphization and subsequent surface passivation are the mechanism for ultralow friction and wear for MCD film. On the other hand, NCD films undergo shear induced plastic deformation without amorphization which leads to the removal of materials during wear test. Hence, the CoF and wear rate increases as a function of decrease in grain size. Based on the observed results, we conclude that MCD films, even with high roughness, show ultralow CoF and lower wear rate. These MCD films find potential utility in mechanical and tribological applications.

**Acknowledgement**

Authors would like to thank Dr. G. Mangamma, Dr S. Dhara and Dr G. Amarendra of the department for their support and encouragement.